\documentclass[useAMS,usenatbib]{mn2e}
\usepackage{epsfig}
\usepackage{rotating}
\usepackage{lscape,graphicx}
\usepackage{color}
\usepackage{natbib}
\usepackage{amsmath}
\usepackage{comment}
\usepackage{color}

\newcommand{\hmsol}{\mbox{$h^{-1}\,{\rm M}_\odot$}}

\providecommand{\kms}{\,\ensuremath{\rm{km\,s}^{-1}}}
\newcommand{\hkpc}{\mbox{$h^{-1}~{\rm kpc}~$}}
\newcommand{\hmpc}{\mbox{$h^{-1}~{\rm Mpc}~$}}

\def\lesssim{\mathrel{\hbox{\rlap{\hbox{\lower3pt\hbox{$\sim$}}}\hbox{\raise2pt\hbox{$<$}}}}}
\def\gtrsim{\mathrel{\hbox{\rlap{\hbox{\lower3pt\hbox{$\sim$}}}\hbox{\raise2pt\hbox{$>$}}}}}

\title[Halo Masses of Mg\,II Absorbers at $z\sim0.5$]{Halo Masses of Mg\,II absorbers at $z\sim0.5$ from SDSS DR7}
\author[J.-R. Gauthier et al. ]{Jean-Ren\'e Gauthier$^{1}$\thanks{E-mail:jrg@astro.caltech.edu (JRG) }, Hsiao-Wen Chen$^{2}$, 
 Kathy L. Cooksey$^{3}$\thanks{NSF Astronomy and Astrophysics Postdoctoral Fellow}, Robert A. Simcoe$^{3,4}$,  
 \newauthor  Eduardo N. Seyffert$^{4}$, and John M. O'Meara$^{5}$  \\
$^{1}$Cahill Center for Astronomy and Astrophysics, California Institute of Technology, Pasadena, CA, 91125, USA\\
$^{2}$Kavli Institute for Cosmological Physics and Department of Astronomy and Astrophysics, University of Chicago, Chicago, IL, 60637, USA \\
$^{3}$MIT Kavli Institute for Astrophysics \& Space Research, Cambridge, MA, 02139, USA \\ 
$^{4}$Department of Physics, MIT, Cambridge, MA 02139, USA \\ 
$^{5}$Department of Chemistry and Physics, Saint Michael's College, Colchester, VT, 05439, USA }
\begin{document}

\date{}

\voffset=-0.6in

\pagerange{\pageref{firstpage}--\pageref{lastpage}} \pubyear{2012}

\maketitle

\label{firstpage}

\begin{abstract}

We present the cross-correlation function of Mg\,II absorbers with respect to a volume-limited sample of luminous red galaxies (LRGs) at $z=0.45-0.60$ using the largest Mg\,II absorber sample and a new LRG sample from SDSS DR7. 
We present the clustering signal of absorbers on projected scales $r_p = 0.3-35$  \hmpc in four $W_r^{\lambda2796}$ bins spanning $W_r^{\lambda2796}=0.4-5.6$\AA. We found that on average Mg\,II absorbers reside in halos $\langle \log M_h \rangle \approx 12.1$, similar to the halo mass of an $L_*$ galaxy. 
We report that the weakest absorbers in our sample with $W_r^{\lambda2796}=0.4-1.1 $\AA\  
reside in relatively massive halos with $\langle \log M_h \rangle \approx 12.5^{+0.6}_{-1.3}$, while 
stronger absorbers reside in halos of similar or lower masses $\langle \log M_h \rangle \approx 11.6^{+0.9}$.
We compared our bias data points, $b$, and the frequency distribution function of absorbers, $f_{W_r}$, with a simple model incorporating an isothermal density profile to mimic the distribution of absorbing gas in halos. We also compared the bias data points with Tinker \& Chen (2008) who developed halo occupation distribution models of Mg\,II absorbers that are constrained by $b$ and $f_{W_r}$. The simple isothermal model can be ruled at a $\approx 2.8\sigma$ level mostly because of 
its inability to reproduce $f_{W_r}$. However, $b$ values are consistent with both models, including TC08. In addition, we show that the mean $b$ of absorbers does not decrease beyond $W_r^{\lambda2796} \approx 1.6$\AA. The flat or potential upturn of $b$ for $W_r^{\lambda2796}\gtrsim 1.6$\AA\ absorbers suggests the presence of additional cool gas in massive halos.
\end{abstract}

\begin{keywords}
galaxies : haloes -- galaxies : quasars absorption lines -- galaxies : general.
\end{keywords}

\section{Introduction}
The relatively long wavelengths and large oscillator strengths of the Mg\,II  absorbers make these systems  
accessible to ground-based spectrographs at $z\gtrsim0.11$, a redshift range where large galaxy samples can 
be effectively collected using 4-m class telescopes. 
Early studies of host candidates of Mg\,II absorbers have confirmed the circumgalactic nature of these 
systems (e.g.,\ \citealt{bergeron1986a,lanzetta1990a,bergeron1991a,bechtold1992a,steidel1994a,bowen1995a}).  
At $z\lesssim 1$, Mg\,II absorbers are found at projected physical separations $\rho \lesssim 100$ kpc of normal galaxies 
characterized by a wide range of colors and luminosities \citep{steidel1994a,chen2008a}. These absorbers 
originate in cool,  photo-ionized gas at temperature $T\sim 10^4$ K and trace high 
column densities of neutral hydrogen N(H\,I)$ \gtrsim 10^{18}$ cm$^{-2}$ \citep{bergeron1986a,churchill2003a,rao2006a}.  
Because of the relatively large oscillator strength, most of the Mg\,II absorbers are likely saturated, especially in SDSS 
spectra. In such cases, the absorption equivalent width reflects the underlying gas kinematics rather than the total gas 
column density. On average, stronger absorbers are found at smaller  
projected distances from the host galaxy (e.g.,\ \citealt{lanzetta1990a,churchill2005a,tripp2005a,kacprzak2008a,barton2009a,chen2010b,nielsen2013a}). 
At the same time, Mg\,II absorbers have also been found to be tracers of galactic-scale outflows in star-forming galaxies 
(e.g.,\ \citealt{weiner2009a,rubin2010a}). 
While several studies have been able to characterize the properties of the CGM based on observations of Mg\,II 
absorption (see \citealt{nielsen2013a}  for a recent compilation), there is still a lack of a general prescription that 
relates these absorbers to the overall galaxy population. Yet it is possible to gain insights into the connection 
between the absorbing gas and luminous matter by measuring their two-point correlation function. 

The spatial two-point correlation function of astrophysical objects is a powerful tool to 
characterize the dark matter halos in which baryons inhabit (e.g.,\ \citealt{davis1983a,zehavi2002a}). 
In recent years, many studies have employed the large-scale clustering signal of various astrophysical objects to 
constrain the mass of the underlying dark matter halos. This technique relies on 
the understanding that the large-scale bias of dark matter halos is monotonically increasing with mass leading to a  
direct relationship between the amplitude of the two-point correlation function and the associated mean halo mass. 
The two-point correlation function of a wide range of astrophysical objects has been studied. 
These objects include, but are not limited to, QSO metal line absorption systems  (e.g., \citealt{adelberger2003a,chen2009a}), 
damped-Ly$\alpha$ systems (e.g., \citealt{bouche2004b,cooke2006a}), low- and intermediate-redshift SDSS galaxies 
(e.g., \citealt{zehavi2002a,tinker2005a}), Quasars (e.g.\ \citealt{ross2009a}), Lyman-break selected 
(e.g.,\ \citealt{bullock2002a,trainor2012a}) and red galaxies at high-$z$ (e.g.,\ \citealt{daddi2003a,hartley2013a}). 

The two-point correlation function of intervening Mg\,II $\lambda\lambda 2796,2803$ absorbers found in QSO spectra was among the early analyses aimed at characterizing the masses of absorber hosts at intermediate redshifts $z\sim0.5$ (e.g.,\ \citealt{bouche2004a}). In Gauthier et al. (2009, hereafter G09), the authors calculated the large-scale two-point cross-correlation function 
between a sample of $\approx 500$ Mg\,II absorbers of strength $W_r^{\lambda2796}>1$\AA\ and a 
\emph{volume-limited} sample of $\approx 200$k luminous red galaxies (LRGs) at $z=0.45-0.60$. 
These authors found a mild decline in the mean halo bias of Mg\,II absorbers with increasing rest-frame absorption 
equivalent width, $W_r(2796)$, similar to the findings of  \citet{bouche2006a} and \citet{lundgren2009a} based on 
a flux-limited sample.
In addition, G09 found a strong clustering signal of Mg\,II absorbers on scales 
$r_p \lesssim 0.3$ \hmpc, indicating the presence of cool gas inside the virial radii of the dark matter halos hosting the passive 
LRGs. This result was later confirmed by the spectroscopic follow-up surveys of \citet{gauthier2010a} and \citet{gauthier2011a}.

The observed $b$--$W_r^{\lambda2796}$ relation has profound implications 
for the physical origin of the Mg\,II absorbing gas. For instance, if more massive halos contain,  on average more Mg\,II gas 
along a given sightine, one would expect a monotonically increasing bias with increasing $W_r^{\lambda2796}$. \citet{tinker2008a} 
developed a halo occupation distribution model and showed that the observed mildly decreasing trend in the $b$--$W_r^{\lambda2796}$ relation is consistent with a transition in the halo gas properties, from primarily cool 
in low mass halos to predominantly hot but with a small fraction of cool gas surviving in high mass halos.

Here we expand upon the G09 analysis, utilizing the largest Mg\,II absorber catalog available from SDSS DR7. 
We extend the 
two-point correlation function to weaker absorbers with $W_r^{\lambda2796}<1$\AA\ while improving 
the precision of the bias measurements of the stronger systems. Specifically, we use a combined Mg\,II catalog 
from \citet{zhu2013a} and \citet{seyffert2013a}. As described in the following section, the combined absorber catalog 
is five times larger than what was used in G09, increasing the statistical significance of the halo bias 
measurements and allowing a clustering analysis based on absorber subsamples from smaller $W_r(2796)$ intervals. 

This paper is organized as follows. In section 2, we present the samples of LRGs and Mg\,II absorbers 
along with the methodology adopted to compute the two-point correlation functions. The 
two-point correlation functions and the derived bias and halo masses of absorber 
hosts are presented in section 3. We discuss the implications of our results for the nature of these absorbers 
in section 4. We adopt a $\Lambda$ cosmology with $\Omega_M = 0.25$ and $\Omega_{\Lambda} = 0.75$ 
throughout the paper. All projected distances are in co-moving units unless otherwise stated and all 
magnitudes are in the AB system. Stellar and halo masses are in units of solar masses. 

\section{Observations and data analysis}
\subsection{LRG catalog} 
The clustering signal of Mg\,II absorbers is computed with respect to a reference population of astrophysical 
objects acting as tracers of the underlying dark 
matter distribution. As discussed in G09, this tracer should be distributed over the 
same imaging footprint as the QSO sighlines that were searched for Mg\,II absorbers. In addition, this reference population 
must have a redshift distribution similar to that of the absorbers. Marked differences in survey mask definition or redshift distributions 
would alter both the shape and the amplitude of the correlation signal in an undesirable fashion. 

Since Mg\,II absorbers can be detected at $z\gtrsim0.35$ in SDSS spectroscopic data, the SDSS LRG sample 
offers the largest reference population yielding a sufficient 
number of galaxy--Mg\,II pairs necessary for a precise calculation of the two-point correlation function. 
LRGs are massive, super-$L_*$ galaxies which are routinely observed from the ground 
with 4-m class telescopes out to $z\sim 0.7$. The clustering signal of LRGs has confirmed 
that these galaxies reside in bias environments characterized by halo masses $\sim 10^{13}$ \hmsol\ 
(e.g., \citealt{padmanabhan2008a,blake2008a}). They are excellent tracers of the large-scale structures in the universe. 

We used the \citet{thomas2011a} photometrically-selected  
LRG catalog (hereafter MegaZDR7) which is based on the SDSS DR7 imaging footprint. 
The catalog comprises 1.4~M entries distributed over 7746 deg$^2$ 
encompassing the spring fields of SDSS located at $\alpha=7^h-19^h$. MegaZDR7 covers 
the photometric redshift range $z_{\rm ph}= 0.38-0.75$. Note that the SEGUE stripes were excluded 
from this catalog. In addition, the three SDSS fall stripes 
(76, 82, 86) were excluded from the survey window function of \citet{thomas2011a} mainly to 
simplify their survey window function. Given the relatively small contribution of these 
stripes, we did not modify the \citet{thomas2011a} galaxy catalog to include these stripes.

\begin{figure}
\centerline{
\includegraphics[angle=0,scale=0.55]{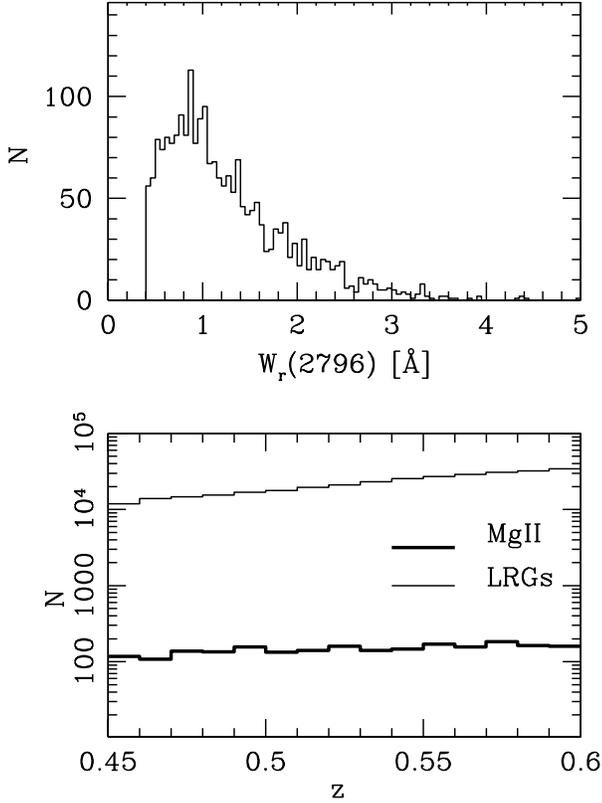}   
}
\caption{\emph{Top:} Rest-frame equivalent width distribution of the 2211 Mg\,II 
absorbers included in the two-point correlation function calculation. \emph{Bottom:} Redshift distribution 
of the volume-limited sample of LRGs (\emph{thin line}) at $z=0.45-0.60$ and Mg\,II absorbers 
(\emph{bold line}).}
\label{distributions}
\end{figure}
 
The LRGs were first selected via a series of cuts in a multidimensional color diagram \citep{collister2007a,blake2008a}
while the photometric redshifts were constructed by an artificial neural network code. In all respects, the 
LRGs were selected in an identical fashion to the galaxy sample employed in G09. As shown in Figure 2 of 
G09, galaxies with $i'>20$ mag have uncertain photo-$z$'s. We thus restricted ourselves to objects 
with $i'_{\rm deV}<20$.\footnote{The $i_{\rm dev}$ magnitudes were corrected for Galactic 
extinction according to the \citet{schlegel1998a} maps.} Typical errors in the photometric redshifts of bright LRGs 
with $i'<20$ are $\sigma_z/(1+z) \approx 0.03 $. Moreover, we applied further cuts by requiring that 
$\delta_{\rm sg}$, the star-galaxy separation parameter, to be $\delta_{\rm sg} > 0.2$. According to 
\citet{collister2007a}, selecting objects with $\delta_{\rm sg}>0.2$ limits the contamination fraction of M 
dwarf stars to $\approx 1.5$\%. These cuts yielded a flux-limited catalog of 1.1M objects.

As discussed in G09, a flux-limited selection criterion creates an inhomogeneous sample of 
LRGs, excluding intrinsically fainter and thus less massive objects at higher redshifts. These authors showed 
that absorber bias may have been overestimated by as much as $\approx 20$\% in previous studies based on 
flux-limited samples of galaxies (e.g.\, \citealt{bouche2006a,lundgren2009a}). Consequently, we adopted 
a volume-limited sample of limiting magnitude 
\begin{equation}
M_{i'} - 5 \log h < -22 
\end{equation}
over the redshift range $z=0.45-0.60$. The limiting magnitude corresponds to the absolute magnitude of our 
faintest galaxies while the redshift range was selected to maximize both the number of LRGs and 
absorbers included in the calculation. Increasing the upper-limit of the redshift range would select intrinsically brighter 
galaxies resulting in a much 
smaller sample size. After excluding all galaxies falling outside our survey mask (see section 2.4.2), our 
final LRG catalog comprises 333k entries, an increase of $\approx 70$\% compared to G09. The photometric 
redshift distribution of the LRGs is presented in the bottom panel of Figure 1.

 \begin{figure}
\centerline{
\includegraphics[angle=0,scale=0.40]{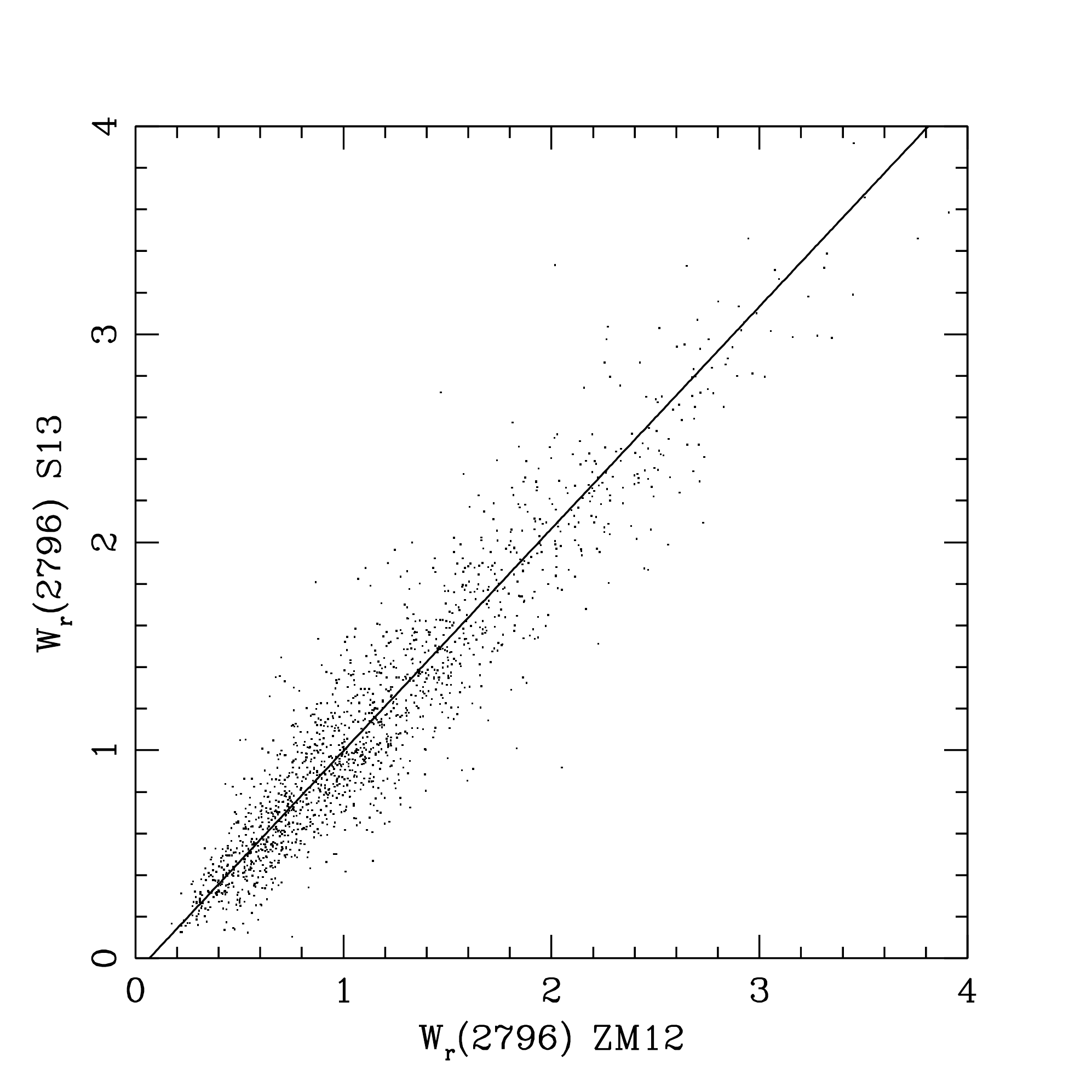}    
}
\caption{Comparison between the rest-frame equivalent width $W_r^{\lambda2796}$ measured by 
both ZM13 and S13 for a subsample of $\sim$ 1500 Mg\,II absorbers listed in both catalogs. 
The solid line corresponds to the best-fit linear regression of the data  
$W_{r,\rm S13}(2796) = 1.07W_{r,\rm{ZM13}}(2796)-0.07$ (\AA).}
\label{ewr_comp}
\end{figure}

\subsection{Mg\,II catalog}
\citet{zhu2013a} and Seyffert et al. (2013) carried out independent searches of Mg\,II 
absorption features in the spectra of distant QSOs found in the SDSS DR7 QSO spectroscopic 
catalog, producing the two largest Mg\,II absorber catalogs that are available in the literature. 
These separate efforts allow us to evaluate the completeness and confirm the accuracy of each 
absorber catalog. By comparing these catalogues, our goal is to establish the most 
\emph{complete} Mg\,II catalog while eliminating as many false positives as possible.

Although both catalogs are derived from essentially the same QSO spectra, we found 
that they significantly  differ in the number of detected absorbers, completeness, and the rate of likely false detections. 
Even when a given absorber is identified by both groups, we found systematic differences in the 
measurements of $W_r^{\lambda2796}$ (see Figure 2). In this section, we describe how we compared and combined these two 
catalogs to yield the final absorber sample adopted for the two-point correlation calculation. 
We first describe each catalog separately along with their respective detection techniques. 
We then discuss the methods we employed to combine the two catalogues 
and produce the final Mg\,II absorber sample. 

\subsubsection{Seyffert et al. (2013) catalog}
The Seyffert et al (2013, hereafter S13) Mg\,II absorber catalog was constructed using
a subset of the SDSS DR7 quasar catalog and
following an equivalent methodology to the C\,IV survey described
in detail in \citet{cooksey2013a}. Here we briefly outline the procedure.

Of the 105,783 quasar spectra in Schneider et al. (2010), 79,595 were
searched for Mg\,II systems if they satisfy the following criteria: (a) The QSOs were not
broad-absorption-line QSOs (i.e. were not listed in \citealt{shen2011a}); (b) The median
signal-to-noise ratio exceeds $\langle {S/N} \rangle = 4$ per pixel in the region 
where intervening absorbers could be detected, outside of the Ly$\alpha$ forest. 

Every quasar spectrum was normalized with a ``hybrid continuum," a fit
combining principle-component analysis, $b$-spline correction, and
pixel/absorption-line rejection. Absorption line candidates were
automatically detected by convolving the normalized flux and error
arrays with a Gaussian kernel with FWHM = 1 pixel, roughly an SDSS
resolution element (resel). The candidate lines with convolved $({
  S/N})_{\rm conv} \ge 3.5$ per resolution element in the $\lambda$2796 line and
$2.5\,{\rm resel}^{-1}$ in $\lambda$2803 were paired into candidate Mg\,II
doublets if the separations of the two lines fell within $\pm 150$ km/s of the 
expected doublet separation $\Delta\,v=767$ km/s.  Any
automatically detected absorption feature with $({S/N})_{\rm conv}
\ge 3.5$ per resolution element and broad enough to enclose a Mg\,II
doublet was included in the candidate list. S13 excluded absorbers 
blueshifted by less than 3000 km/s from the QSO redshift. For the purposes of this 
constraint, the quasar redshifts were taken from Schneider et al. (2010), 
not from \citet{hewett2010a}, as was used in \citet{zhu2013a} and the  
rest of the current work.

All candidates were visually inspected by at least one author 
of S13 and most by two. They were rated on a four-point scale from
0 (definitely false) to 3 (definitely true). The systems were judged
largely on the basis of the expected properties of the Mg\,II doublet (e.g., centroid alignment,
correlated profiles) but also including possible, associated ions for verification. Any 
system with rating of 2 or 3 were included in subsequent analyses.

The wavelength bounds of the absorption lines were automatically
defined by where the convolved $S/N$ array began increasing when
stepping away from the automatically detected line centroid. The new
centroid was then set to be the flux-weighted mean wavelength within
the bounds, and the Mg\,II doublet redshift was defined by the
new centroid of the $\lambda$2796 transition. The sum of the absorbed flux within the bounds sets
the equivalent width. In summary, the S13 catalog contains 35,629 absorbers 
over the redshift range $z=0.4-2.3$. 

\subsubsection{Zhu \& Menard (2013) catalog} 
We also examined the recently published Zhu \& M\'enard (2013, hereafter ZM13) Mg\,II catalog. The ZM13
catalog is based on a sample of 85,533 QSO sightlines distributed over the SDSS DR7 
Legacy and SEGUE spectroscopic footprints. In addition, ZM13 included 1411 QSOs from 
the \citet{hewett2010a} sample that were not identified in Schneider et al. (2010). Their catalog consists of 35,752 intervening 
Mg\,II absorbers over the redshift range $z=0.4-2.3$. 

In brief, ZM13 used a non-negative principle component analysis (PCA) 
and a set of eigenspectra to estimate the continuum level of each QSO spectrum. Further improvements 
on the continuum estimation was done by applying two median filters of 141 and 71 pixels in width. This process 
was repeated three times until convergence was achieved. 

Once the continuum is determined, candidate Mg\,II 
absorbers were selected via a matching filter search involving the Mg\,II doublet and four Fe\,II lines. Absorbers 
were identified if their $S/N$ was above a minimum threshold of 4 for the $\lambda2796$ line and 2 for the $\lambda2803$ line. 
Each candidate doublet was then fitted with double-Gaussian profiles and candidates were rejected if the measured doublet separation 
exceeded 1\AA. To further eliminate false positives, ZM13 made use of the Fe\,II $\lambda2586$ and $\lambda2600$ lines to measure the $S/N$ 
of the four lines and applied a cut on the $S/N$ to reject false positives. Finally, $W_r^{\lambda2796}$ and $z_{\rm Mg\,II}$ were 
determined by fitting a Gaussian (or double Gaussian) to the candidate profile. In summary, the ZM13 method is fully automated and 
involves little human intervention. 

 \begin{figure}
\centerline{
\includegraphics[angle=0,scale=0.44]{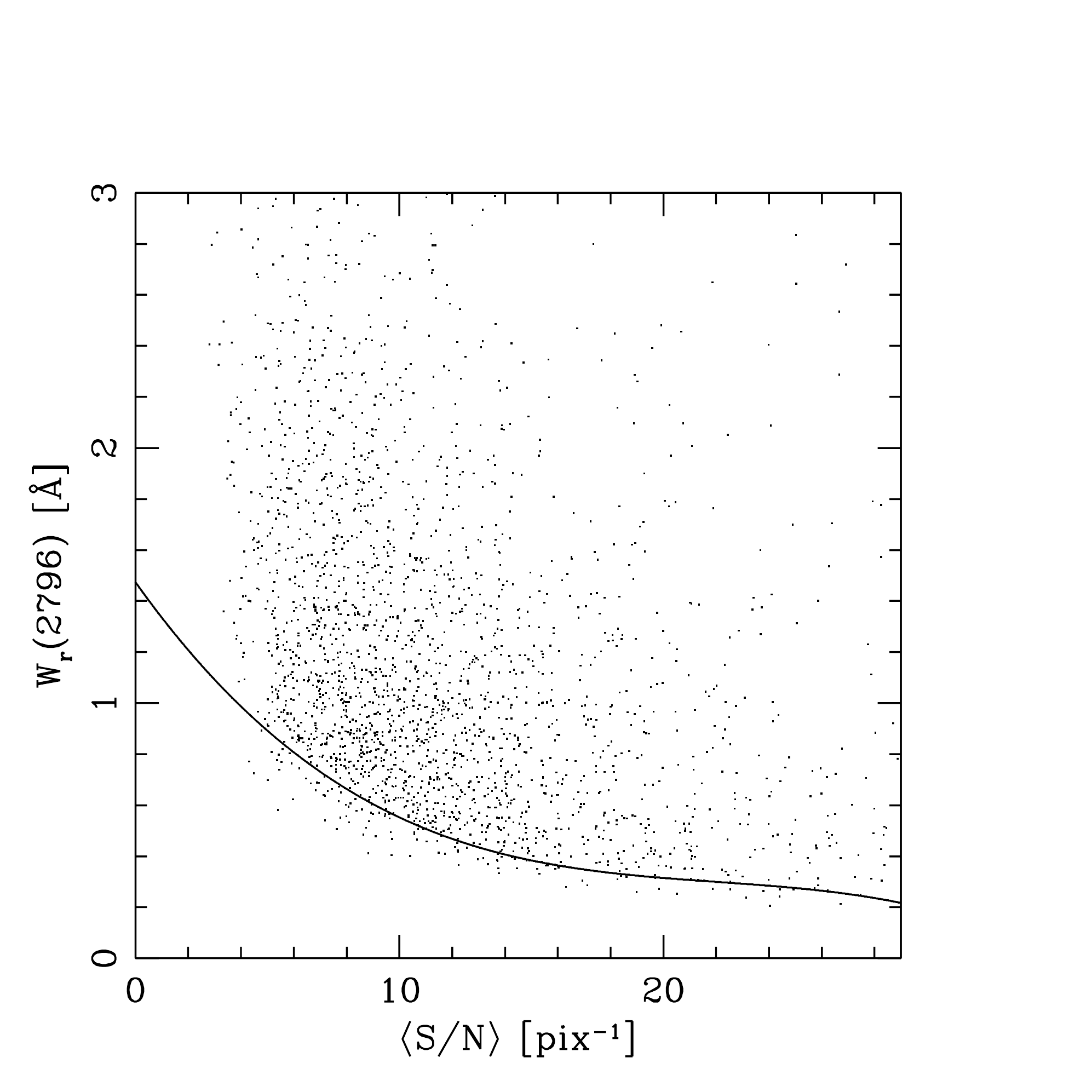}    
}
\caption{The rest-frame equivalent width $W_r^{\lambda2796}$ of Mg\,II absorbers as measured by S13 with respect to the median 
$S/N$ per pixel, $\langle S/N \rangle$, of the QSO spectra over the redshift range $z_{\rm Mg\,II}=0.45-0.60$. 
The solid line is a 3rd-order polynomial fit to the bottom 5th percentile of the $W_r^{\lambda2796}$ distribution. We interpret 
this fit as  the typical lower-limit on $W_r^{\lambda2796}$ that can 
be measured along a QSO sightline of given $\langle S/N \rangle$. The determination of this lower limit on $W_{r}(S/N)$  
is necessary to generate the catalog of random absorbers (see section 2.3.4). 
 }
\label{snr_g_wr}
\end{figure}

\subsubsection{Comparing and combining the Mg\,II catalogs}

We first compared the two catalogs and identified common absorbers by selecting those with the same 
RA and DEC coordinates. In addition, we made sure that for each matched absorber, the observed wavelength 
listed in one catalog's was falling within the wavelength bounds of the other and vice versa. 
In the redshift range $z_{\rm Mg\,II}=0.45-0.60$, at velocity separation $\delta v>10000$ \kms\ 
below the QSO redshift and within our survey mask (see section 2.3.2) we found 1491 matched 
absorbers in the ZM13 and S13 catalogs. The redshifts of these absorbers, as measured by S13 
is consistent with the published values of ZM13. We adopted the redshifts 
of S13. 

However, we found that $W_r^{\lambda2796}$ measured by ZM13 is systematically 
lower than S13. In Figure \ref{ewr_comp}, we show a comparison of $W_r^{\lambda2796}$ measured 
for $\sim 1500$ matched absorbers randomly selected from both catalogs. In this Figure, the solid line corresponds to the best-fit linear regression 
of $W_r^{\lambda2796}$ S13 with respect to $W_r^{\lambda2796}$ ZM13.  We found 
\begin{equation}
W_{r,\rm{S13}}(2796) = (1.07\pm 0.01) \times W_{r,\rm{ZM13}}(2796) - (0.07\pm0.01)
\end{equation} 
where $W_{r,\rm{S13}}(2796)$ is measured by S13 and $W_{r,\rm{ZM13}}(2796)$ by ZM13.  

To determine which measurement of $W_r^{\lambda2796}$ should be adopted in our final absorber catalog, we examined 
$\approx$ 100 randomly selected absorbers in ZM13 and S13.  We first fitted the QSO spectra with a series 
of third-order $b-$splines to set the continuum level in the spectral regions of the absorbers and we visually 
established the boundaries of the absorbing region. We determined $W_r^{\lambda2796}$ by integrating the 
flux decrement between the absorption boundaries. Our measurements are in agreement with S13. We found 
that ZM13 tends to underestimate the continuum level, although the differences are typically $\lesssim0.1$\AA. We thus adopted 
the $W_r^{\lambda2796}$ estimates of S13 and we applied the above $W_r^{\lambda2796}$ correction for absorbers found 
only by ZM13. 

To establish our final absorber catalogs, we applied further selection criteria. In addition to the absorber 
redshift range, QSO--absorber velocity separation, and survey footprint cuts discussed above, we also applied 
cuts to eliminate false detections. We first applied a series of $W_r$ ratios to eliminate systems with either 
very large  ($W_r^{\lambda2796}/W_r^{\lambda2803}>2.2$) or very low ($W_r^{\lambda2796}/W_r^{\lambda2803}<0.59$)
column density ratios that are inconsistent with Mg\,II absorbers being either optically thin or completely saturated. 
In addition, we applied a combined 2.9$\sigma$ detection threshold on the doublet components decomposed into a 2.5$\sigma$ ($1.5\sigma$)
threshold on the $\lambda2796$ ($\lambda2803$) transition. Furthermore, one of us (KLC) visually inspected 
all the remaining systems and further eliminated likely false positives. Finally, we eliminated four absorbers 
found by ZM13 that occur in the Hewett \& Wild (2010) sample of 1411 visually-inspected QSO sightlines because there is no 
$S/N$ estimate readily available for these QSO sightlines (see section 2.3.4). 

These selection criteria yielded a sample 2323 absorbers. We further restricted the sample to absorbers 
with $W_r^{\lambda 2796}>0.4$\AA\ to ensure a large enough sample ($\approx 10000$) of QSO sightlines with 
sufficient $S/N$ to detect the \emph{weakest} absorbers. This selection criterion is particularly 
important when generating the random absorber catalog for the weakest systems (see section 2.3.4). This final 
cut reduces the number of absorbers to 2211. 

Of these 2211 absorbers, 1260 were found by ZM13 and S13, 
793 were only found by S13, and 158 only by ZM13. A number of reasons explain why 793 absorbers were found 
by S13 and not by ZM13. Among them, 265 absorbers were found within $|\delta z| = 0.02$ redward of the 
QSO C\,IV emission with the QSO emission redshift defined by Hewett \& Wild (2010). In addition, 40 absorbers 
were found within  $|\delta z| = 0.04$ blueward of the QSO Mg\,II emission, and 8 were found in near Galactic 
Ca\,II H\&K absorbers.  A combination of factors, including differences in the automatic line detection 
algorithm and user biases could potentially explain why the remaining 480 systems were missed by ZM13 
although estimating the relative importance of each factor is beyond the scope of this paper (see S13 for further 
details). 

As for the 158 absorbers detected by ZM13 and not by S13, 6 occur in sightlines with $\langle S/N \rangle< 4$ pix$^{-1}$ in 
the region searched for Mg\,II absorbers, 53 are found in sightlines identified as BALs by \citet{shen2011a}, and 99 have too low $\langle S/N \rangle_{\rm conv}$ 
in either  the $\lambda 2796$ or $\lambda2803$ spectral regions to be automatically detected by S13. 

In summary, our final catalog  consists of $2211$ absorbers with 
$W_r^{\lambda2796}=0.40-5.6$\AA\ and distributed over the redshift range $z=0.45-0.60$.  Note 
that the typical redshift error of these absorbers is very small and of order $10-20$km/s. 
In Figure \ref{distributions}, we show the redshift distributions of LRGs and Mg\,II absorbers 
as well as the $W_r^{\lambda2796}$ distribution. The distribution of $z_{\rm Mg\,II}$ is flat while the photo-$z$ distribution increases 
sightly toward higher $z$.  We divided this Mg\,II sample according to $W_r^{\lambda2796}$ into four bins of approximately equal 
number of absorbers. Each bin contains roughly the same number of absorbers as the entire Mg\,II sample considered 
by G09. The bins considered have $W_r=[0.4-0.78],[0.78-1.08],[1.08-1.59]$, and $[1.59-5.6]$ \AA. 

\subsection{Two-point correlation statistics} 

\subsubsection{Method}

A detailed description of the method to measure the two-point correlation function is presented 
in G09. In summary, we adopted the Landy \& Szalay (1993, hereafter LS93) minimum variance 
estimator to calculate the projected two-point auto- and cross-correlation functions ($w_p$) 
between Mg\,II absorbers and LRGs. The LS93 estimator is 
\begin{equation}
w_p(r_p) = \frac{D_aD_g-D_aR_g - D_gR_a +R_aR_g}{R_aR_g}
\label{ls93}
\end{equation}
where $D$ and $R$ are data and random points; $a$ and $g$ refer to absorbers
and galaxies; and $r_p$ is the projected co-moving separation on the sky between two objects.  
We adopted the same binning as in G09 and divided the pairs into eight $r_p$ bins equally 
spaced in logarithmic space and covering the range $0.2-35$ \hmpc. The upper limit of 35 
\hmpc is a few times smaller than the size of our jackknife cells (see section 2.4.3). In the 
following subsections, we discuss the selection of a survey mask and the methodology 
adopted to generate random galaxies $R_g$ and absorbers $R_a$. 

\subsubsection{Survey mask} 
When computing the $w_p(r_p)$ statistics, both data and randoms should be distributed on 
the same survey mask. If galaxies and absorbers occupy survey windows that are not 
completely overlapping on the sky, the shape and amplitude of the correlation signal 
would be altered in an undesirable fashion. Hence it is crucial to identify a survey mask that is 
common to both LRGs and Mg\,II absorbers and large enough to minimize the number of 
objects falling outside the survey mask. In addition, 
the same survey mask shall be used to distribute random LRGs and Mg\,II absorbers. 

We adopted the  SDSS DR7 spectroscopic angular selection function mask\footnote{The file used was $\rm{sdssdr72safe0res6d.pol}$ 
available 
at http://space.mit.edu/~molly/mangle/download/data.html.} 
provided by the NYU Value-Added Galaxy Catalog team \citep{blanton2005a} and assembled 
with the Mangle 2.1 software \citep{hamilton2004a,swanson2008a}. This mask 
represents the completeness of the SDSS spectroscopic survey as a function of the 
angular position on the sky.  Since LRGs were not identified by \citet{thomas2011a} in 
stripes 76,82 and 86, we also excluded these stripes from the survey window. 
In addition, our adopted survey mask does not include the SEGUE spectroscopic mask. 

\begin{figure*}
\centerline{
\includegraphics[angle=0,scale=0.77]{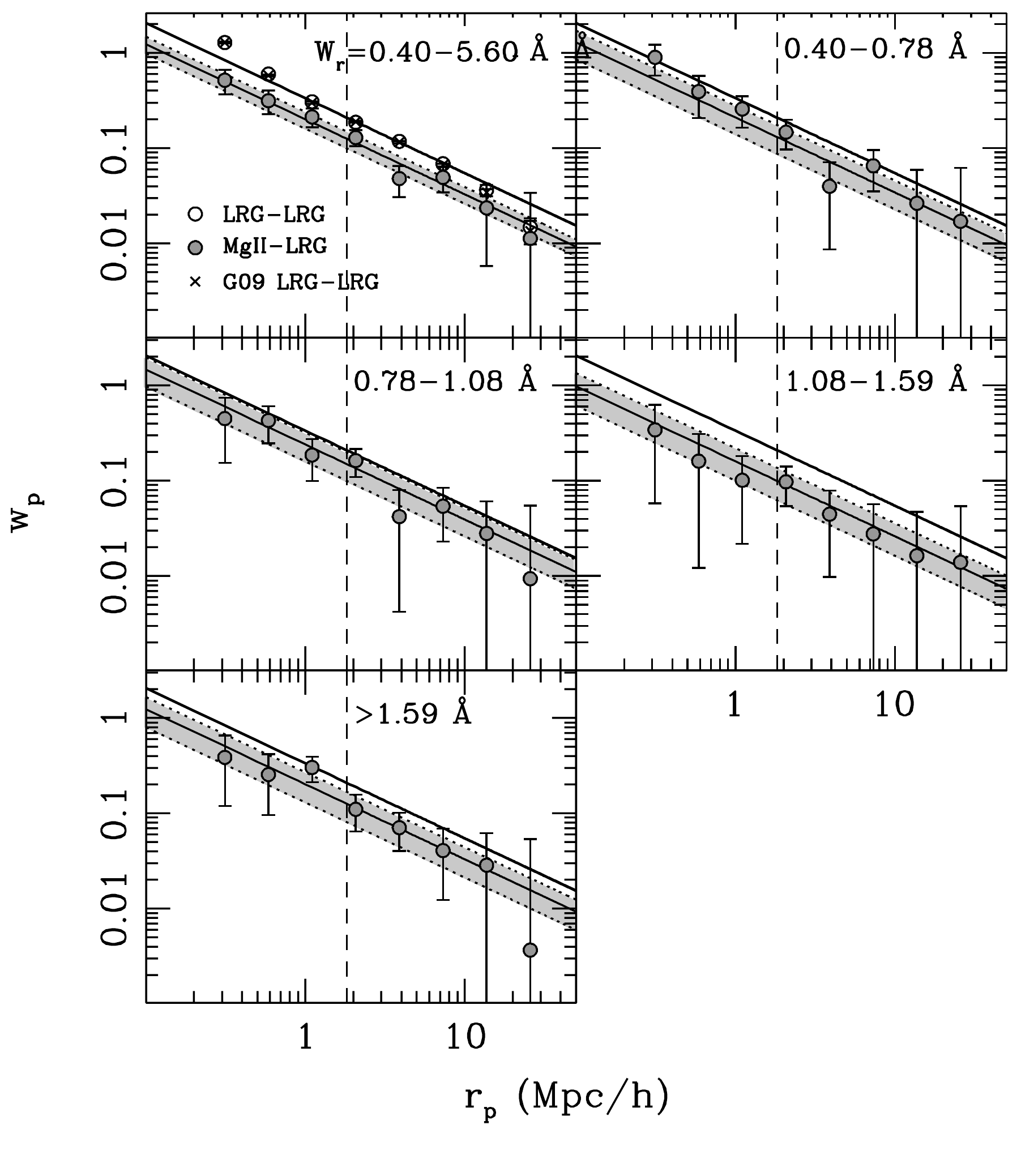}   
}
\caption{Two-point cross- and auto-correlation functions for the volume-limited sample of LRGs 
at $z=0.45-0.60$. In the \emph{top left} panel, the LRG auto-correlation signal is represented by 
\emph{open circles} while the Mg\,II--LRG cross-correlation is shown in \emph{solid grey circles}. 
For comparison, we show the LRG--LRG auto-correlation function from G09 in \emph{crosses}. 
To guide the eye, we included a thick solid line corresponding to the best-fit power-law model $f(x) = ax^b$ of
 the LRG auto-correlation signal at $r_p \gtrsim 2$ \hmpc. The vertical \emph{dashed} lines demarcate the small-scale (one-halo) 
$w_p$ from the large-scale (two-halo) $w_p$ on which the bias calculation is based. In this case, $b=-0.787$ and $a=0.335$. To facilitate the 
 comparison between the clustering amplitude of absorbers and LRGs, we repeated this solid curve in all five remaining panels. 
 The \emph{thin black} curve corresponds to the best-fit power-law 
 model of the cross-correlation function in which we fixed the power-law slope to the value derived 
 for the LRG auto-correlation signal ($b=-0.787$). The dotted lines and and shaded areas correspond to the 1$-\sigma$ 
 error bars on the amplitude $a$.  
The Mg\,II absorbers are divided in five bins according to the rest-frame equivalent 
width of the $\lambda2796$ transition $W_r^{\lambda2796}$. Each panel corresponds 
to a different $W_r^{\lambda2796}$ bin labelled in the upper right corner. No correction for photometric 
redshifts is applied in this figure.  The correction will be applied when estimating 
the relative bias of Mg\,II absorber hosts (see section 3.2). G09 showed that uncertainties in photometric redshifts of the LRGs 
decreases the clustering amplitude ratio between Mg\,II absorbers and the LRGs. } 
\label{xx}
\end{figure*}

\subsubsection{The errors on $w_p$}
We consider two independent sources contributing to the errors in the measured $w_p$:   
cosmic variance and photometric redshift uncertainties. The contribution of cosmic variance can be estimated 
by using the jackknife resampling technique applied to the survey mask. The 
sky was separated into $N=192$ cells of similar survey area ($\approx 40$ deg$^2$). The choice of the
number of cells was obtained after running convergence 
tests on a varying number of cells (see G09 for more details). The cosmic variance estimate for each $r_{p,i}$ 
corresponds to the $i$th-diagonal element of the covariance matrix calculated with 
the jackknife method, 
\begin{equation}
{\rm COV}(w_i,w_j)=\frac{N-1}{N}\sum_{k=1}^{N}(w_i^k-\overline{w_i})(w_j^k-\overline{w_j})
\label{covariance}
\end{equation}
where $k$ represents the iteration in which box $k$ was removed from the calculation. 
The mean value $\bar{w_i}$ was calculated for bin $i$ over all $w_p^k$'s. Although $r_p$ 
bins are correlated on large scales we consider only the diagonal elements of the covariance 
matrix when estimating the contribution of cosmic variance to the error on $w_p$. 

The second source of errors comes from large uncertainties in the photometric redshifts 
of the galaxies. As discussed in G09, photometric redshift uncertainties affect $w_p$ in two distinct ways. 
The first effect is to \emph{lower} the overall clustering amplitude by introducing uncorrelated 
pairs in the calculation (see section 3.2). The second effect is to add random noise in the 
calculation. This is particularly important for the inner $r_p$ bins which particularly suffer from 
an uncertainty on $z_{\rm ph}$ which translates into a large fractional uncertainty on $r_p$. 

To estimate the random noise in $w_p$ as a result of photo-$z$ uncertainties, we followed 
the procedure described in G09. In brief, for each cross- and auto-correlation calculation, 
we generated 100 independent realizations of the LRG catalog by sampling the photo-$z$ 
distribution of each galaxy. We considered that each photo-$z$ distribution can be characterized 
by a normal distribution with $\sigma_z=0.03(1+z_{\rm ph})$ and a mean value $z_{\rm ph}$ \citep{collister2007a}. 
We then calculated $w_p$ for each realization and determined the photo-$z$ contribution 
to the error on $w_p$ by calculating the dispersion among these 100 realizations for each 
$r_p$ bin. 

We found that while photometric redshift uncertainties can increase the random error by 
$\lesssim 20$\% at $r_p\lesssim 1$\hmpc, the effect is negligible for larger values of $r_p$. Typically 
at these larger $r_p$ values, the increase is $<3$\% for the LRG auto-correlation signal. Not surprisingly, the effect is smaller 
for the cross-correlation calculation since only one of the two pair members has photometric redshift. 
These findings are consistent with G09. Since the contribution of 
photo-$z$'s uncertainties is negligible on large scale where the clustering amplitude is calculated, 
we thus only considered cosmic variance in the error budget of $w_p$. Note however that photo-$z$'s 
also lowers the amplitude of the clustering signal and this effect is significant and is discussed 
in section 3.2.

\subsubsection{Generating random LRGs and Mg\,II absorbers} 
The RA and DEC positions of each random LRG was generated using the 
\emph{ransack} routine available through the Mangle software package. 
The redshifts of the random galaxies were determined by sampling the 
redshift distribution of the LRGs (see Figure 1). The number of random galaxies was 
determined after running convergence tests. As discussed in G09, 
having $\sim$ 10 times more random galaxies than the actual number of 
LRGs is sufficient to achieve convergence. Consequently, we generated 
a catalog of 4M random galaxies. 
 
We randomly assigned the the RA and DEC positions of the random absorbers  
among the QSO sightlines that have been surveyed by ZM13 and S13 and fell within our 
survey mask.  Of these sightlines, 63,294 were 
found within our spectroscopic survey mask. The redshift of each random 
Mg\,II was determined by sampling the redshift distribution of the absorber 
data while $W_r^{\lambda2796}$ was assigned by sampling the distribution 
$dN/W_r \propto \exp(-W_r/W_*)$ where $W_*$ is the typical absorber strength 
given by equation (5) in ZM13. 

To allocate a given random absorber to a QSO sightline, one has to consider 
two additional limitations. First, the absorber should be found at velocity
separation $|\delta_v|>10,000$\kms\ from the QSO redshift and outside of the Ly$\alpha$ 
forest. This effectively
eliminates all QSO sightlines with $z_{\rm QSO}<0.45$ and $z_{\rm QSO}>2.67$. 
Second, the QSO spectra should have high enough $S/N$ 
to detect a random absorber of a given strength.  
This constraint becomes particularly important when generating 
random absorbers for the \rm{weak} absorber bin ($W_r^{\lambda2796}=0.40-0.78$\AA). 

To establish whether or not a given sightline 
could harbor a random absorber of strength $W_r^{\lambda2796}$, we empirically determined 
the minimum $W_r^{\lambda2796}$ ($W_{r,\rm min}$) that could be detected at a given $S/N$ 
by adopting the following procedure. First, we determined the median $S/N$ per pixel, 
$\langle S/N \rangle$,  over the redshift range $z_{\rm Mg\,II}=0.45-0.60$ for all QSO 
spectra in the Schneider et al. (2010) catalog. Next, we compared the strength of the 
Mg\,II absorbers listed in S13 with $\langle S/N \rangle$. The results are shown in Figure 
\ref{snr_g_wr}. We computed a 3rd-order polynomial fit to the bottom 5th-percentile 
of the $W_r^{\lambda2796}$ distribution and interpret this fit as the minimum absorber strength, $W_{r,\rm min}(S/N)$,
that could be detected given $\langle S/N \rangle$.  The fit is 
shown by the solid line in Figure \ref{snr_g_wr}. Note that the value of $W_{r,\rm min}(S/N)$ corresponds 
approximately to a 2.5-$\sigma$ detection across the whole range of $\langle S/N \rangle$ considered. 
A random absorber of strength $W_r^{\lambda2796}>W_{r,\rm min}$  
could be detected and is included in our random absorber catalog. 
We repeated the procedures described above until we collected a sample 
of 100K random absorbers.  We adopted this number of random absorbers 
after performing convergence tests, as described in G09. 

\section{Results} 
In this section, we present the projected two-point 
cross-correlation functions LRG--Mg\,II along with the LRG 
auto-correlation function for a volume-limited sample of LRGs at $z=0.45-0.60$. 
We explored the dependence of the clustering signal on $W_r^{\lambda2796}$ 
by computing the LRG--Mg\,II cross-correlation for five $W_r^{\lambda2796}$ bins
with each bin having a similar number of absorbers. A summary of each 
calculation, including the adopted $W_r^{\lambda2796}$ binning, the number of 
absorbers and LRGs along with the number of $D_aD_g$ pairs found in the 
first $r_p$ bin is presented in Table 1. 

In Figure \ref{xx}, we show the auto- and cross-correlation functions. In the top left panel, we show 
the LRG--LRG auto-correlation function in \emph{open circles}. The error bars on the auto-correlation 
function are typically smaller than the size of the symbol. We calculated 
the best-fit power law model $f(x)=ax^b$ of the auto-correlation signal at $r_p\geq $ 2 \hmpc\ 
and displayed the result as the thick solid line. We limited the fit to data points in the two-halo regime where 
the bias is determined. As discussed in Zehavi et al. (2004), correlation functions of galaxies display departures 
from power law models, especially at large separations. In the LRG auto-correlation function, this departure 
is more pronounced at $r_p>10$ \hmpc\ and demonstrates the limitation of using such power law fits to 
determine the bias of galaxies and absorber hosts. In \emph{grey circles}, we show the LRG--Mg\,II cross-correlation 
signal. We also adopted a power-law model for the cross-correlation function, but we fixed the slope to the 
value obtained for the LRGs ($b=-0.787$). The thin black line and the shaded areas correspond to the 
best-fit and 1-$\sigma$ errors on the amplitude $a$. Note that the best-fit power-law models are only shown 
to guide the eye and do not enter in the calculation of the relative bias of absorber hosts. 
The other five panels display the cross-correlation signals for the five $W_r^{\lambda2796}$ bins 
considered. The range of $W_r^{\lambda2796}$ for each bin is labeled in the upper-right corner 
of each panel. In all panels,  the error bars on $w_p$ correspond to cosmic variance, estimated 
using the jackknife resampling technique (see section 2.4.3). Note that the data 
points shown in Figure \ref{xx} have \emph{not} been corrected for the systematic effects of 
photometric redshifts on the clustering amplitude. These considerations 
will enter in the calculation of the relative bias of absorber hosts and are discussed 
in section 3.2. 

\begin{figure*}
\centerline{
\includegraphics[angle=0,scale=0.70]{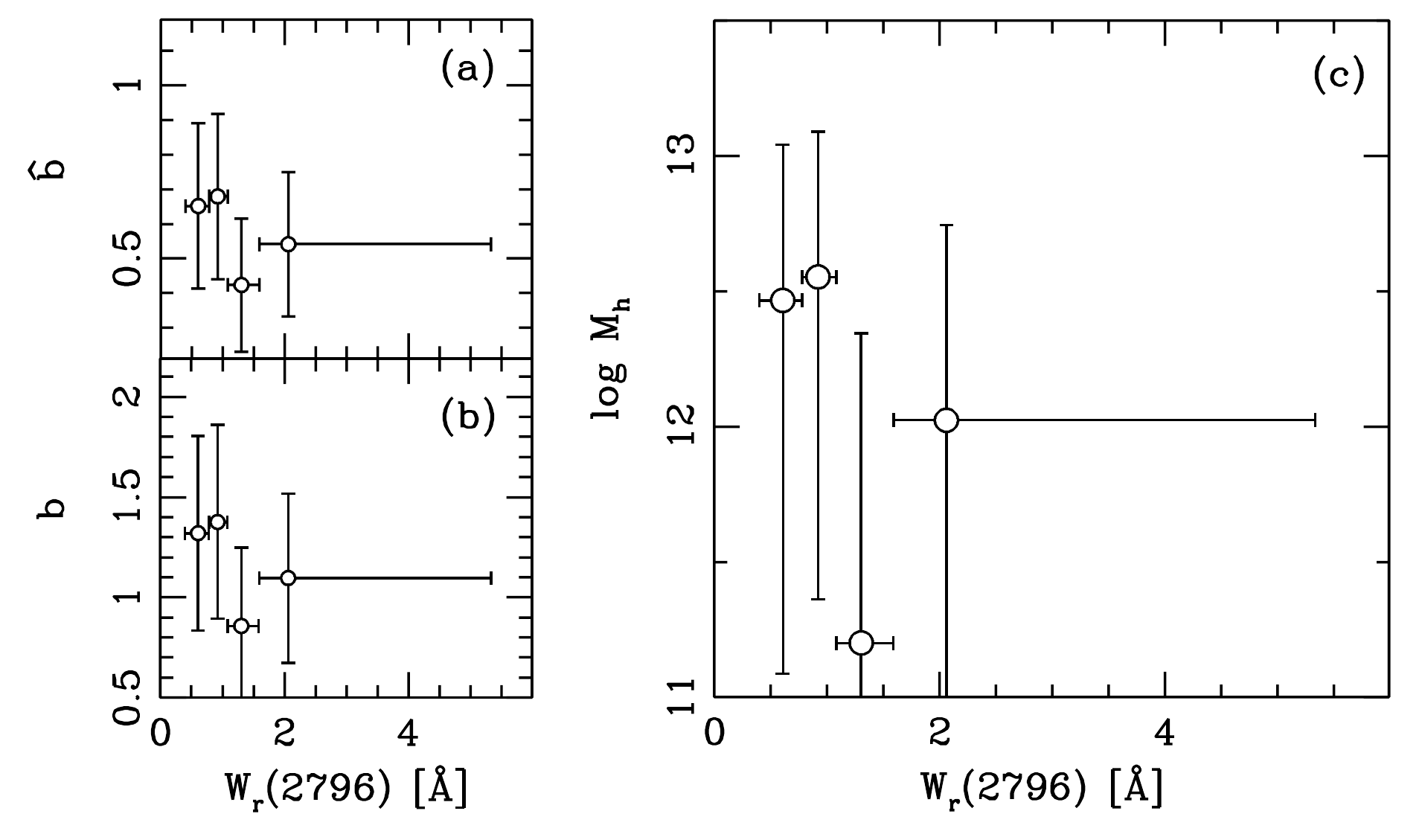}  
}
\caption{(\emph{a}) Relative bias of Mg\,II hosts derived from the direct ratio method for points at $r_p\geq 2$\hmpc. 
We show the results for the four $W_r^{\lambda2796}$ bins considered 
in this analysis. (\emph{b}) Absolute bias of Mg\,II absorber hosts. To compute the bias, we adopted the bias of 
LRGs $b_g=2.023 \pm 0.006$, as measured in G09.
(\emph{c}) In \emph{open} circles, we show halo masses derived by inverting the $b(M)$ relationship. Note 
that the lower error-bar on four 
of the five $W_r^{\lambda2796}$ bins are arbitrary. This occurs when the lower error bar reaches
 $b \approx 0.7$ giving us no constraint on the 
halo mass. The points represent the median of the $W_r^{\lambda2796}$ distribution 
in each bin and the error bars on $W_r^{\lambda 2796}$ represent the extent of each bin. 
}
\label{rel_bias}
\end{figure*}

At $r_p\gtrsim$ 2 \hmpc, where LRG--LRG and Mg\,II--LRG pairs are probing distinct dark 
matter halos (i.e.,\ in the two-halo regime), the weakest absorbers 
($W_r^{\lambda2796} < 1.08$\AA) have the largest cross-correlation amplitude 
similar to the LRG auto-correlation signal. 
In contrast, absorbers in the bin $W_r^{\lambda2796}=1.08-1.59$\AA\ show, on average, the lowest 
clustering amplitude on large scales. 

\subsection{Theoretical framework on bias and halo mass determination}
Here we provide a brief discussion of the theoretical framework behind the determination of the 
Mg\,II absorber host bias and halo mass from the projected two-point correlation function. A more detailed 
overview can be found in \citet{berlind2002a}, \citet{zheng2004a}, and \citet{tinker2005a}. 

The bias of dark matter halos of mass $M$ is the ratio of the real-space two-point correlation function of these 
dark matter halos, $\xi_h(M,r)$, and the dark matter auto-correlation function $\xi_m( r)$ 
\begin{equation}
b_h^2(M,r)=\frac{\xi_h(M,r)}{\xi_m(r)} \; . 
\end{equation} 
In practice, we compute the projected two-point correlation statistics by integrating the real-space 
correlation function along the line-of-sight ($\pi$) direction 
\begin{equation}
w_p(r_p) = \int_l \xi(r_p,\pi)d\pi \; . 
\end{equation}
The LRG--LRG auto-correlation signal $\xi_{gg}$ can be decomposed into a mass dependent $b_g^2(M_g)$ 
and scale dependent $f_g^2( r)$ term 
\begin{equation}
\xi_{gg}(M_g, r) = b_g^2(M_g) f_g^2( r) \xi_m( r)
\end{equation}
where $M_g$ is the bias-weighted mean halo mass. Similarly, the cross-correlation absorber--galaxy term can be written as  
\begin{equation}
\xi_{ga}(M_g,M_a,r) = b_g(M_g) b_a(M_a) f_g( r) f_a( r)\xi_m( r) ,
\end{equation}
where the terms with subscript $a$ refer to the absorbers. 
On large scales and for the halo masses considered in this paper, the scale dependence of $b_h^2(M,r)$, $f( r)$, is 
almost independent of halo mass and is divided out when computing the relative bias 
of absorber hosts, $\hat{b}$. Consequently,  one can write 
\begin{equation}
\hat{b} \equiv \frac{b_a(M_a)}{b_g(M_g)} = \frac{\xi_{ga}( r)}{\xi_{gg}( r)} = \frac{w_{ag}(r_p)}{w_{gg}(r_p)} 
\end{equation}
where $\hat{b}$ is the scale-independent relative bias of absorber hosts $a$ with respect to a population of galaxies $g$. 
In other words, $\hat{b}$ can be calculated from the ratio of the projected two-point correlation functions $w$ on 
large scales. From $\hat{b}$, the bias (and mass) of the absorber hosts can be derived if the absolute bias of 
the tracer galaxy population ($b_g$) is known. 
For this purpose, we adopted the same galaxy bias as in G09. Since the LRG selection criteria and 
redshift range are the same as G09, the auto-correlation signal of LRGs should, in principle 
be the same. We verified that this was the case. In the upper-left panel of Figure \ref{xx}, we show the LRG 
auto-correlation signal of G09 in \emph{crosses}. As expected,  the differences from 
G09 are negligible. We thus adopted the bias value of $b_{g}=2.023\pm0.006$ as found by G09. 

\subsection{Calculating the relative bias }

Before calculating the relative bias of Mg\,II absorber hosts, it is important to consider the effects of 
photometric redshifts on the clustering amplitude and bias measurements. LRGs typically have 
photo-$z$ accuracy of $\sigma_z\approx 0.03(1+z_{\rm ph})$. Large uncertainties on the galaxy 
redshifts affect the clustering signal in two different ways. Since the redshifts are uncertain, the 
angular diameter distance is uncertain too, introducing uncertainties in the projected co-moving 
separations between LRGs and Mg\,II absorbers and effectively ``smoothing" out sharp features present in the 
intrinsic two-point correlation signal. As discussed in section 2.4.3, this source of uncertainty mostly affects 
the inner bins of the correlation function and is negligible at $r_p \gtrsim$ 2 \hmpc where the clustering 
amplitude is measured. 

Furthermore, photometric redshift errors effectively \emph{broadens} the redshift range included in the 
calculation and adds uncorrelated galaxy--Mg\,II pairs in the calculation. This effect results in a 
systematic lowering of the \emph{amplitude} of the clustering signal. G09 addressed this issue by calculating 
the projected clustering signal on mock LRG distributions. Their mock catalog was produced by 
populating dark matter halos of an $N-$body simulation with a halo occupation distribution function 
determined from a spectroscopic LRG sample. Then the redshift of each galaxy was  perturbed to mimic 
the effects of photometric redshifts and the authors re-calculated the two-point correlation statistics. 
In a nutshell, G09 found that the MgII--LRG cross-correlation 
amplitude on large scales ($>1$ \hmpc) is reduced by a factor 0.79$\pm$0.02 compared to what is found for 
a spectroscopic sample of LRGs while the LRG--LRG auto-correlation amplitude is reduced by 0.71$\pm$0.01. 
Consequently, we determined the relative bias, $\hat{b}$, by multiplying the uncorrected $\hat{b}$ 
values by the correction factor $\mathcal{C}=0.90\pm0.02$.  Hereafter, all $\hat{b}$ and absolute bias measurements $b$ 
are corrected for this systematic effect introduced by photometric redshifts.  
 
In G09, they discussed the direct ratio (DR) method to calculate the relative bias of absorber hosts. This technique 
employs the weighted mean ratio of all data points at $r_p \gtrsim 2$ \hmpc\ to obtain $\hat{b}$ 
\begin{equation}
\hat{b}=\sum_{i=4}^{8} \omega_i \frac{w_{ag,i}}{w_{gg,i}}  \times \mathcal{C}
\end{equation}
where the weights, $\omega_i$ are given by 
\begin{equation}
\omega_i = \frac{w_{ag,i}}{\sigma_i^2 w_{gg,i}} \; . 
\end{equation}
The index $i$ denotes the $r_p$ bin and $\sigma_i$ is the error on $w_{ag,i}/w_{gg,i}$ using 
error propagation technique as is the error on $\hat{b}$ (see G09). In Table 2, we provide estimates for the relative 
bias $\hat{b}$, bias $b$, and halo masses for each one of the five $W_r^{\lambda2796}$ bins 
considered and the entire Mg\,II sample. 

\subsection{Absolute bias and halo mass}
As discussed in section 3.1, the halo mass of absorber hosts could be obtained simply by inverting 
the $b(M)$ relationship. We refer to this method as \emph{bias-inverted} halo mass. 
Moreover, one can estimate halo masses by integrating the bias-weighted 
halo mass function down in mass until the bias value reaches $b$. The minimum 
halo mass corresponding to $b$ can thus be used as a lower-limit in the integral of the 
halo mass function. This \emph{bias-weighted} halo mass estimate is simply derived by integrating the 
mass-weighted halo mass function using this lower-limit. 

These two techniques were discussed in details in G09. The authors show that the methods give halo mass estimates 
within 0.1 dex of each other for both LRGs and Mg\,II absorber hosts. Consequently, we followed the G09 methodology 
and used the bias-inverted masses as our Mg\,II hosts mass estimates. 

In Figure \ref{rel_bias}, we show the relative bias, bias, and halo masses of Mg\,II absorber hosts for the 
five $W_r^{\lambda2796}$ bins considered. Except for the weakest absorbers, 
the lower error bars we quoted on the halo masses are arbitrary. This lack of constraints arises because when the lower error 
bar on the bias reaches 0.7, there is no constraint on the halo mass. In fact, $b(M)$ reaches a minimum value 
of $\approx0.7$ and becomes nearly independent of halo mass at log $M_h \lesssim 9$ (e.g.,\ \citealt{tinker2008b}). 

\begin{figure}
\centerline{
\includegraphics[angle=0,scale=0.70]{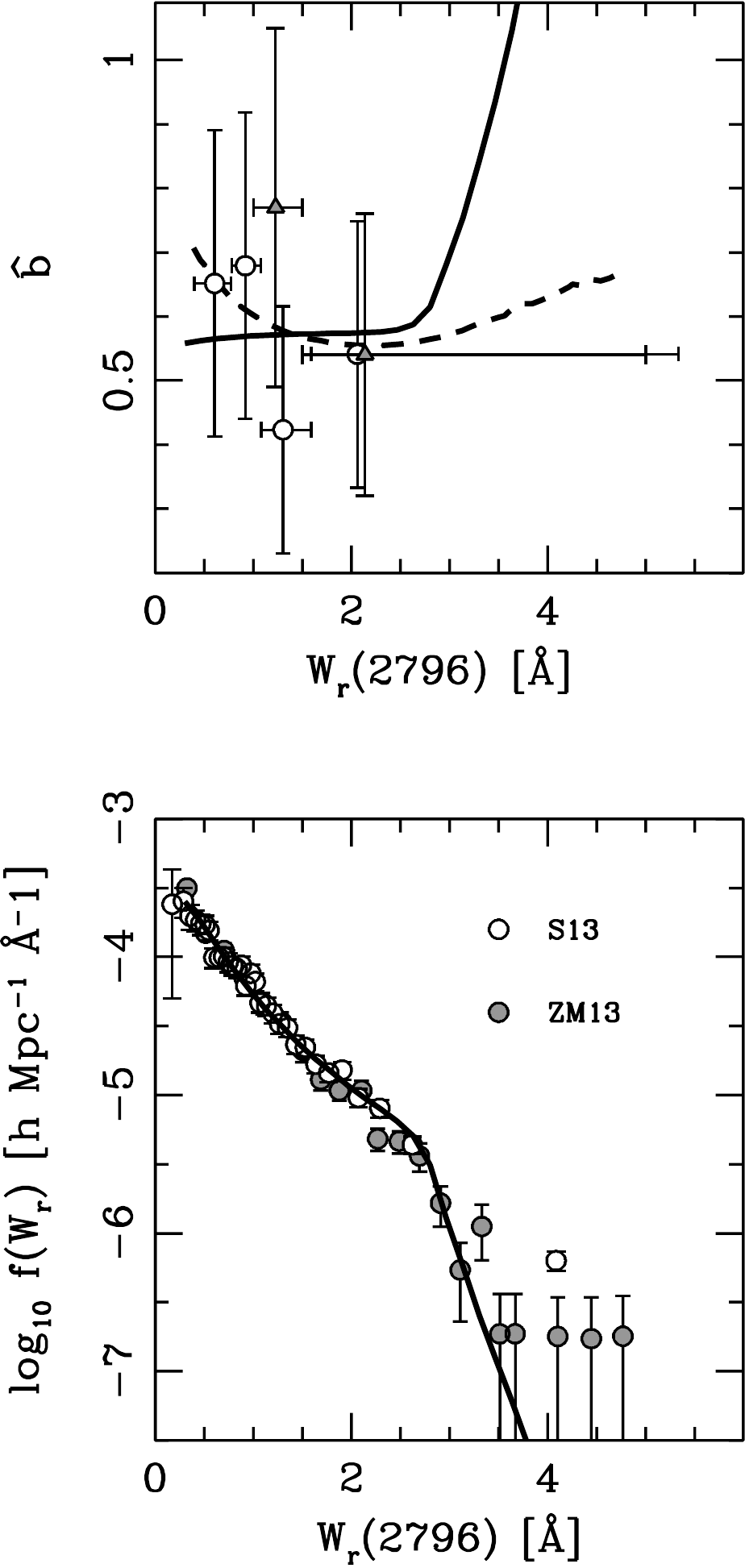}    
}
\caption{ \emph{Top:} Comparison between 
the relative bias estimates of Mg\,II absorber hosts and
the predictions from a simple model in which the gas distribution 
follows an isothermal profile (see section 4 for further details). In \emph{open circles} are the $\hat{b}$ 
values found in this paper. In \emph{grey triangles} we show the results 
from G09.  
The \emph{solid} line corresponds to our best-fit model.  Interestingly, this model  predicts a 
monotonically increasing $b$--$W_r$ relation. We also present a direct comparison between the predictions of the 
``transition" model from Tinker \& Chen (2008) (TC08) in \emph{dashed} curve 
(see section 4 for further details).  This transition models predicts an anti-correlation 
between $b$ and $W_r$ and is in better agreement with our data, especially for the 
weakest $W_r$ bins. 
\emph{Bottom:} Comparison between the best-fit model and the 
frequency distribution function of absorbers, $f(W_r)$, measured by S13 (\emph{open circles}) in the redshift range $z_{\rm Mg\,II}=0.45-0.60$. 
We also plotted the ZM13 $f(W_r)$ values in \emph{grey} circles over the range $z_{\rm Mg\,II}=0.43-0.55$ for comparison.  
 Because of the small $f(W_r)$ error bars for points at $W_r\lesssim2$\AA, the $\chi^2$ is driven by the weak 
 $W_r$ bins. The model tends to underestimate $f(W_r)$ for strong absorbers although a larger covering fraction for 
 massive halos or a higher $A_{W_0}$ would over predict the number of weaker systems while flattening the overall $f(W_r)$ slope.  
 For 32 degrees of freedom, this model has a reduced $\chi^2$, $\chi_r^2 = 1.81$ which has an associated $P$-value $P=0.003$.
 Accordingly, this simple model can be rejected at a $\approx 2.8\sigma$ (one-sided) confidence level. 
}
\label{rel_bias_models}
\end{figure}

\begin{table}
 \centering
 \begin{minipage}{140mm}
  \caption{Description of the cross-correlation calculations}
  \begin{tabular}{@{}cccc@{}}
  \hline
   $W_r^{\lambda2796}$ [\AA]  & N$_{\rm Mg\,II}$ & N$_{\rm LRGs}$ & DD pairs (1st bin)  \\   
   \hline
  All & 2211 &  333334 & 130  \\
  $0.40-0.78$ & 559 & 333334 & 36 \\ 
  $0.78-1.08$ & 531 & 333334 & 26\\
  $1.08-1.59$ & 564 & 333334 & 25 \\
  $>1.59$ & 557 & 333334 & 24 \\
 \hline
\hline
\end{tabular}
\end{minipage}
\end{table}

\section{Discussion}
We presented the cross-correlation function of Mg\,II absorbers with respect to a volume-limited sample 
of LRGs at $z=0.45-0.60$ using the SDSS DR7 galaxy and absorber catalogs. We benefited from 
an absorber catalog four times larger and an increase of 70\% in the number of LRGs compared 
to the samples used in G09. We extended the clustering analysis to weaker absorbers with $W_r^{\lambda2796}<1$\AA\  
that were excluded from the G09 analysis. 

The clustering signal of Mg\,II absorbers was calculated in four $W_r^{\lambda2796}$ bins of roughly equal 
number of absorbers spanning the range $W_r^{\lambda2796}=0.4-5.6$\AA.  On average, stronger absorbers 
with $W_r^{\lambda2796}>1$\AA\ reside in less massive halos than weaker ones. 

\begin{table}
 \begin{minipage}{80mm}
  \caption{Relative bias, absolute bias, and halo mass estimates of Mg\,II absorber hosts}
  \begin{tabular}{@{}ccccccccccc@{}}
  \hline
   $W_r^{\lambda2796}$ [\AA]  & $\hat{b}$\footnote{We applied a correction factor of 0.90$\pm$0.02 
   that takes into account the broader redshift interval introduced by photo-$z$, as discussed in G09}  & $b$\footnote{We adopted the bias of LRGs $b_g=2.023\pm0.006$ as 
   measured in G09.} & $\langle \log M_h \rangle $\footnote{The halo mass corresponds to the \emph{bias-inverted} halo mass, i.e. we invert the $b$ vs $M_h$ relationship 
   to find the halo mass corresponding to the measured bias.}   \\   
   \hline

   All  &  0.56 $\pm$ 0.12   & 1.14 $\pm$ 0.23 & $12.1^{+0.4}_{-0.7}$ \\
   $0.4-0.78$ &  0.65 $\pm$ 0.24 &  1.32 $\pm$ 0.48 & 12.5$^{+0.6}_{-1.4}$\\
   $0.78-1.08$   & 0.68 $\pm$ 0.24 &  1.38 $\pm$ 0.48 & 12.6$^{+0.5}_{-1.2}$\\
   $1.08-1.59$  & 0.42 $\pm$ 0.19 & 0.86 $\pm$ 0.39 &  11.2$^{+1.1}$\\ 
   $>1.59$ & 0.54 $\pm$ 0.21 &  1.10 $\pm$ 0.42 &  12.0$^{+0.7}$ \\

 \hline
\hline
\end{tabular}
\end{minipage}
\end{table}

The observed $b$--$W_r^{\lambda2796}$ relation offers a statistical characterization of the origin of the Mg\,II absorbers as 
a whole, but the interpretation requires a comprehensive model of halo gas content as a function of halo mass. Similarly, 
the frequency distribution function $f(W_r)$ of absorbers, namely the number of absorbers per unit $W_r$ per unit comoving length, 
also provides important constraints on halo gas models. We thus compared these two datasets with the predictions of a simple analytical 
model in which the Mg\,II gas is distributed in dark matter halos according to an isothermal profile. In this model, the gas clumps are distributed according to an isothermal 
profile of finite radial extent. This  radius is denoted by $R_{\rm gas}$, the gaseous radius of the halo. In our model, the value of 
$R_{\rm gas}$ is set to $R_{\rm gas} =1/3 \times R_{\rm vir}$ where $R_{\rm vir}$ is the virial radius of the dark matter 
halo following \citet{chen2010a}. For example, a typical $L_*$ galaxy with halo mass $\sim 10^{12}$ \hmsol\ at $z\sim 0.25$ has $R_{\rm gas}=75$ \hkpc \citep{chen2010a}.  
The total absorption equivalent width is the sum over all clumps encountered along the sightline 
and thus corresponds to the integral of the isothermal profile along a given sightline located at impact parameter $s$ 
\begin{equation}
W_r(s,M) = A_W\frac{2\mathcal{G}_0}{\sqrt{s^2+a_h^2}} \arctan \sqrt{\frac{R_{\rm gas}^2-s^2}{s^2+a_h^2}}
\label{wr_model}
\end{equation}
where $a_h$ is the core radius of the isothermal profile, $A_W$ is the mean absorption equivalent width per 
unit surface mass density of the cold gas and $\mathcal{G}_0$ is 
\begin{equation}
\mathcal{G}_0 = \frac{M(<R_{\rm gas})/4\pi}{R_{\rm gas}-a_h \arctan (R_{\rm gas}/a_h)} \; . 
\end{equation} 
Following Tinker \& Chen (2008, hereafter TC08), we set $a_h=0.2R_{\rm gas}$ and adopted a power-law for $A_W$ 
\begin{equation}
A_W(M)= A_{W_0} \bigg( \frac{M}{10^{12} h^{-1} M_{\odot} }\bigg)^{-0.2}~[h^{-1} \rm{\AA} ~\rm{cm}^{2}~\rm{g}^{-1} ]
\end{equation}
where $A_{W_0}$ is a free parameter independent of halo mass. The slope of this power-law corresponds to the 
best-fit value of TC08. The frequency distribution function can be written as 
\begin{equation}
f(W_r) \equiv \frac{d^2N}{dW_r dl} = \int dM \frac{dn}{dM} \sigma_g(M) P(W_r|M) 
\end{equation}
where $dn/dM$ is the halo mass function (Warren et al. 2006), $\sigma_g(M)=\pi R_{\rm gas}^2$ is the gas cross section and 
$P(W_r|M)$ is the probability that a halo of mass $M$ hosts an absorber of strength $W_r$. In turn, $P(W_r|M)$ can 
be written as : 
\begin{equation}
P(W_r|M) = \kappa_g(M) \frac{2s(W_r|M)}{R_{\rm gas}^2} \frac{ds}{dW_r} 
\label{prob}
\end{equation}
(see equation 10 of TC08). In this expression, the impact parameter $s(W_r|M)$ is found by inverting equation 
\ref{wr_model} which is performed numerically. The derivative of $s$ with respect to 
$dW_r$, $ds/dW_r$,  is calculated following equation 11 in TC08.  In equation \ref{prob},  $\kappa_g(M)$ is the gas 
covering fraction. We adopted a double power law to account for the mass dependence of $\kappa$ 
\begin{equation}
\kappa(M) = \begin{cases} \kappa_{12}(M/10^{12})^{\gamma}, & \mbox{if } M \leq 10^{12} \hmsol \\ \kappa_{12}(M/10^{12})^{\alpha}, & \mbox{if } M>10^{12} \hmsol \end{cases}
\end{equation}
Following the empirical findings 
of \citet{chen2010a},  we assigned $\kappa_{12}=0.7$ which corresponds to the covering fraction of $W_r^{\lambda 2796}>0.3$ \AA\ 
absorbers in $L_*$-galaxy halos.  Both slopes $\alpha$ and $\gamma$ are free parameters.

In addition to $f(W_r)$, the model described above allows us to predict the $b$--$W_r$ relation. 
The bias corresponds to the mean 
halo bias weighted by the probability of finding an absorber $W_r$ in a halo of mass $M$ 
\begin{equation}
b=\frac{1}{f(W_r)} \int dM \frac{dn}{dM} \sigma_g(M) b_h(M) P(W_r|M)
\end{equation}
where $b_h$ is the halo bias taken from \citet{tinker2008b}. 

In summary, this model has three free parameters: $A_{W_0}$, $\alpha$, and $\gamma$. We generated a grid of 
models by varying these three parameters independently and we compared our model predictions for $f(W_r)$ and $b$ with the 
empirical data of S13 in the range $z_{\rm Mg\,II}=0.45-0.60$ and the bias data points of Figure 5a. In addition, we plotted in \emph{grey}
the ZM13 $f(W_r)$ data over the redshift range $z_{\rm Mg\,II}=0.43-0.55$ for comparison. 
For each model, we computed a $\chi^2$ goodness-of-fit test \begin{equation}
\chi^2 = \chi_b^2 + \chi_f^2 = \sum_{i=0}^4 \frac{(b_i-\bar{b}_i)^2}{\sigma_{b_i}^2} + \sum_{j=0}^{31} \frac{(f_j - \bar{f}_j)^2}{\sigma_{f_j}^2}
\end{equation}
where $\chi_b^2$ and $\chi_f^2$ correspond to the values from the bias and frequency distribution function. The parameters 
$\bar{b}$ and $\bar{f}$ are the model predictions for the bias and frequency distribution function respectively. 
Because of the larger number of datapoints of the frequency data (31 vs 4), $\chi_f^2$ will dominate over $\chi_b^2$ for 
most values of $(A_{W_0},\alpha,\gamma)$. The best-fit model has $\chi^2=57.9$ with $\chi_b^2=0.96$ and $\chi_f^2=56.97$ 
The corresponding best-fit parameters are $(126,-1.40,2.18)$. For a $10^{13} \hmsol$ halo, this model predicts $\kappa=0.03$, 
which is lower than the value obtained by \citet{gauthier2011a}  for LRG halos ($\kappa_{\rm LRG}=0.22\pm0.13$). We show the best-fit model in Figure 6. Because 
of the small $f(W_r)$ error bars for points at $W_r\lesssim2$\AA, $\chi_f^2$ is driven by the weak $W_r$ bins. The model tends to 
underestimate $f(W_r)$ for strong absorbers although a larger $\kappa$ for massive halo or a larger $A_{W_0}$ would over predict
the number of weaker systems while flattening the overall $f(W_r)$ slope.  Interestingly, this model  
predicts a monotonically increasing $b$--$W_r$ relation. For 32 degrees of freedom, the reduced 
$\chi^2$ is $\chi_r^2 = 1.81$ which has an associated $P$-value of $P\approx0.003$. Accordingly, this simple model can be rejected at a 
$\approx 2.8\sigma$ (one-sided) confidence level. 

In contrast with the simple model presented above, TC08
also  developed a halo occupation distribution model (HOD) in which they introduced a transition mass 
scale of $\sim 3\times10^{11}$ \hmsol\ above which a shock develops and reduce the amount of Mg\,II absorbing 
gas within the shock radius.  This transition mass scale was motivated by recent hydrodynamical simulations 
of galaxy formation showing two distinct channels of gas accretion for halos above and below the 
transition scale (e.g. \citealt{keres2009a}), although the conclusions drawn from these early studies 
are now being challenged by recent hydrodynamical simulations (e.g.,\ \citealt{nelson2013a}). 
TC08 showed that their model predicts near unity covering fraction of Mg\,II absorbing gas over a wide 
range of halo masses, suggesting that Mg\,II absorbers are probing an unbiased sample of galaxies, 
not preferentially selected for their recent star formation activity.
Although this ``transition" model provides a better fit to $f(W_r)$, the most important differences occur 
in the $b$--$W_r$ relationship. Because of the shock radius developing in massive halos, the transition model 
predicts a suppressed contribution from massive halos yielding a lower $b$ for stronger absorbers. This 
results in an overall $b$--$W_r$ anti-correlation. In Figure \ref{rel_bias_models} we show a direct comparison 
between the relative bias and the model predictions taken from TC08.  
The best-fit TC08 ``transition" model is shown in \emph{dashed} and provides a better fit to the bias data although 
most of the discriminative power lies in the weakest $W_r$ bin. 

The bias data shown in Figure 5 suggest a possible flattening or upturn in the $b$--$W_r$ relation for absorbers 
with $W_r^{\lambda2796}\gtrsim 1.59$\AA . A similar,  trend was also seen in \citet{bouche2006a} and \citet{lundgren2009a}. 
These results are consistent with \citet{gauthier2013a} who argue that ultra-strong Mg\,II absorbers with 
$W_r^{\lambda2796}\gtrsim 3$\AA\ trace gas dynamics of the intragroup medium. Gauthier (2013) 
estimated halo masses of $\log M_h=12-13.3$ for the galaxy groups presented in their paper and in \citet{nestor2011a}. 

Similar group environments have also been found around $W_r^{\lambda2796}\approx 2$\AA\  
absorbers at intermediate redshifts \citep{whiting2006a,kacprzak2010b}. 

Nevertheless, the SDSS-III Baryon Oscillation Spectroscopic Survey (BOSS) is collecting a large sample of spectroscopically 
identified LRGs with a redshift precision of $dz<0.0005$ instead of $dz\sim0.05$.  The availablity of a spectroscopic 
LRG sample offers an exciting opportunity to study the real-space clustering signal of Mg\,II absorbers and to examine the 
two-point function on small scales ($<$1Mpc).  These new measurements will provide further insights into large-scale 
motion of cool gas uncovered by Mg\,II and allow a detailed investigation of the cool halo gas content of massive halos hosting the LRGs.

\section*{Acknowledgments}
It is a pleasure to thank Michael Rauch and Jeremy Tinker for helpful comments and discussions. 
JRG gratefully acknowledges the financial support of a Millikan Fellowship provided by Caltech. 

The SDSS MgII catalog (Seyffert et al. 2013) was funded largely by the National Science 
Foundation Astronomy \& Astrophysics Postdoctoral Fellowship (AST-1003139) and in part by 
the MIT Undergraduate Research Opportunity Program (UROP) Direct Funding, from the 
Office of Undergraduate Advising and Academic Programming and the John Reed UROP Fund.

We are grateful to the SDSS collaboration for producing and maintaining the SDSS public data archive. 
Funding for the SDSS and SDSS-II has been provided by the Alfred P. Sloan Foundation, the Participating Institutions, 
 the National Science Foundation, the U.S. Department of Energy, the National Aeronautics and Space Administration, 
the Japanese Monbukagakusho, the Max Planck Society, and the Higher Education Funding Council for England. 
The SDSS Web Site is http://www.sdss.org/.

The SDSS is managed by the Astrophysical Research Consortium for the Participating Institutions. The Participating 
Institutions are the American Museum of Natural History, Astrophysical Institute Potsdam, University of Basel, 
University of Cambridge, Case Western Reserve University, University of Chicago, Drexel University, Fermilab, 
the Institute for Advanced Study, the Japan Participation Group, Johns Hopkins University, the Joint Institute 
for Nuclear Astrophysics, the Kavli Institute for Particle Astrophysics and Cosmology, the Korean Scientist Group, 
the Chinese Academy of Sciences (LAMOST), Los Alamos National Laboratory, the Max-Planck-Institute for Astronomy (MPIA), 
the Max-Planck-Institute for Astrophysics (MPA), New Mexico State University, Ohio State University, 
University of Pittsburgh, University of Portsmouth, Princeton University, the United States Naval Observatory, 
and the University of Washington.

\footnotesize{
\bibliographystyle{mn2e}
\bibliography{ms}
}

\appendix

\label{lastpage}

\end{document}